\documentclass[longbibliography,aps,prb,twocolumn,superscriptaddress,showkeys,amsfonts,amssymb,amsmath,floatfix,verbatim]{revtex4-2}
\usepackage{inputenc}
\usepackage{amsmath,amsbsy,amsfonts,revsymb}
\usepackage{graphicx}
\usepackage{bm}
\usepackage[dvipsnames]{xcolor}
\usepackage[separate-uncertainty=true,multi-part-units=single]{siunitx}
\usepackage{bm}
\usepackage{mathtools} 
\usepackage[colorlinks=true,linkcolor=blue,citecolor=blue,urlcolor=blue,breaklinks=true]{hyperref}
\usepackage{soul}

\DeclarePairedDelimiter\dpa{(}{)} 

\DeclarePairedDelimiter\dca{[}{]}

\DeclarePairedDelimiterX\Bra[1]{\langle}{\rvert}{#1\,}

\DeclarePairedDelimiterX\Ket[1]{\lvert}{\rangle}{\,#1}

\DeclarePairedDelimiterX\Abs[1]{\lvert}{\rvert}{\,#1\,}
\DeclarePairedDelimiterX\qsp[2]{\langle}{\rangle}{#1\delimsize\vert\mathopen{}#2}
\DeclarePairedDelimiterX\Qsp[2]{\langle}{\rangle}{#1\,\delimsize\vert\,\mathopen{}#2}
\DeclarePairedDelimiterX\mel[3]{\langle}{\rangle}{#1\delimsize\vert\mathopen{}#2\delimsize\vert\mathopen{}#3}
\DeclarePairedDelimiterX\Mel[3]{\langle}{\rangle}{#1\,\delimsize\vert\,\mathopen{}#2\,\delimsize\vert\,\mathopen{}#3}

\newcommand\mrm[1]{\mathrm{#1}}

\begin{document}
	
\title{Vibronic collapse of ordered quadrupolar ice in the pyrochlore magnet Tb$_{2+x}$Ti$_{2-x}$O$_{7+y}$}

\author {Y. Alexanian}
\email[]{yann.alexanian@unige.ch} 
\altaffiliation[]{Present address: Department of Quantum Matter Physics, University of Geneva, 24 Quai Ernest-Ansermet, 1211 Geneva, Switzerland}
\affiliation{Universit\'e Grenoble Alpes, CNRS, Institut N\'eel, 38000 Grenoble, France}

\author{J. Robert}
\affiliation{Universit\'e Grenoble Alpes, CNRS, Institut N\'eel, 38000 Grenoble, France}

\author{V. Simonet}
\affiliation{Universit\'e Grenoble Alpes, CNRS, Institut N\'eel, 38000 Grenoble, France}

\author {B. Langérôme}
\affiliation{Synchrotron SOLEIL, L’Orme des Merisiers, 91192 Gif-sur-Yvette, France}

\author {J.-B. Brubach}
\affiliation{Synchrotron SOLEIL, L’Orme des Merisiers, 91192 Gif-sur-Yvette, France}

\author {P. Roy}
\affiliation{Synchrotron SOLEIL, L’Orme des Merisiers, 91192 Gif-sur-Yvette, France}

\author{C. Decorse}
\affiliation{ ICMMO, Universit\'e Paris-Saclay, CNRS, 91400 Orsay, France}

\author{E. Lhotel}
\affiliation{Universit\'e Grenoble Alpes, CNRS, Institut N\'eel, 38000 Grenoble, France}

\author{E. Constable}
\affiliation{Institute of Solid State Physics, TU Wien, 1040 Vienna, Austria}

\author {R. Ballou}
\affiliation{Universit\'e Grenoble Alpes, CNRS, Institut N\'eel, 38000 Grenoble, France}

\author {S. De Brion}
\email[]{sophie.debrion@neel.cnrs.fr}
\affiliation{Universit\'e Grenoble Alpes, CNRS, Institut N\'eel, 38000 Grenoble, France}

\date{\today}
	
\begin{abstract}

While the spin-liquid state in the frustrated pyrochlore Tb$_{2+x}$Ti$_{2-x}$O$_{7+y}$  has been studied both experimentally and theoretically for more than two decades, a definite description of this unconventional state still needs to be achieved. Using synchrotron-based THz spectroscopy in combination with quantum numerical simulations, we highlight a significant link between two features: the existence of a quadrupolar order following an ice rule and the presence of strong magneto-elastic coupling in the form of hybridized Tb$^{3+}$ crystal-field and phonon modes. The magnitude of this so-called vibronic process, which involves quadrupolar degrees of freedom, is significantly dependent on small off-stoichiometry $x$ and favors all-in-all-out like correlations between quadrupoles. This mechanism competes with the long range ordered quadrupolar ice, and for slightly different stoichiometry, is able to destabilize it.

\end{abstract}

\maketitle

\section{Introduction} 


Quantum materials have attracted considerable interest lately. A wide spectrum of exotic behaviors built upon magnetic dipolar degrees of freedom were already discovered and are still foreseen. Among them, one finds those induced through magnetic frustration such as quantum spin liquids with fractional or anyonic excitations to quote but one example \cite{lacroix2010,Kitaev2006}. Another emblematic geometric frustration-induced state is the classical spin ice, as observed in Ho$_2$Ti$_2$O$_7$ \cite{Harris1997}, belonging to the pyrochlore family $R_{2}M_{2}$O$_{7}$, where the magnetic rare-earth ions $R^{3+}$ form a corner-sharing tetrahedra network \cite{Gardner2010}. Spin ice correlations are characterized by the ice rule (two spins pointing into and two out of each tetrahedron), which can be destabilized towards an exotic ordered/disordered fragmented state when additional competing all-in-all-out correlations (all spins pointing into or out of each tetrahedron) are present \cite{BrooksBartlett2014,Lefrancois2017,Cathelin2020}. A novel field of unconventional behaviors induced by frustration has recently emerged, involving multipolar degrees of freedom that come into play once the angular momenta are larger than one half. A case in point is provided by the rare-earth-based compounds \cite{Santini2009}, the description of which sometimes require Hamiltonians involving quadrupoles \cite{Morin1990,Araki2012,Tsunetsugu2021} or even multipoles of higher orders, such as recently shown in Ce-, Pr- or Nd-based pyrochlores \cite{Onoda2010,Onoda2011,Huang2014,Petit2016,Benton2020,Sibille2020,Xu2020,Leger2021}. These multipoles are governed by the nature of the magnetic ion as well as their local environment, the so called crystal field that lifts the degeneracy of their electronic levels. Extensions of theses degrees of freedom to composite hidden order \cite{Spaldin2020} were also considered that might be relevant to quite different quantum materials such as high temperature superconductors \cite{Lovesey2015,Fechner2016}. Here, we focus on quadrupolar degrees of freedom and show how they play a crucial role in the exotic and still enigmatic behavior of  the pyrochlore compound Tb$_{2}$Ti$_{2}$O$_{7}$.\nocite{Taniguchi2013,Wakita2016}
First studies on Tb$_{2}$Ti$_{2}$O$_{7}$ more than two decades ago showed that no long-range magnetic order is stabilized down to at least 70~mK \cite{Gardner1999,Gardner2001}. The compound remains in an intriguing spin liquid state quite different from conventional spin ices, although similarities have been observed \cite{Fennell2012,Petit2012}. While it was quickly noticed that the presence of a Tb$^{3+}$ excited crystal field level as low as $\Delta =$ 1.5~meV (12~cm$^{-1}$) \cite{Gingras2000,Gardner2001,Mirebeau2007,Lummen2008,Bertin2012,Zhang2014,Klekovkina2014,Princep2015,Ruminy2016b} certainly has a central role in this exotic quantum state \cite{Kao2003,Enjalran2004}, any tentative description of the fundamental state failed to reproduce all experimental data \cite{Molavian2007,Curnoe2008,Molavian2009,Bonville2011,Petit2012,Curnoe2013,Bonville2014}. In the past decade, two important steps were achieved almost simultaneously. On one hand, it was realized (based on specific heat \cite{Taniguchi2013,Wakita2016} and latter supported by elastic constants \cite{Gritsenko2020} measurements), that Tb$_{2+x}$Ti$_{2-x}$O$_{7+y}$ has a strong sensitivity to a very low anti-site Tb/Ti disorder. For the off-stoichiometry parameter $x\gtrsim 0$, it enters into a quadrupolar long range ordered phase below $T_{\mrm{c}}\approx$ 0.5~K \cite{Kadowaki2015,Kermarrec2015,Takatsu2016,Kadowaki2018,Kadowaki2019,Kadowaki2022}. On the other hand, while no static distortion is observed \cite{Ruff2007,Lummen2008,Nakanishi2011,Goto2012,DalmasDeReotier2014,Ruminy2016c}, extensive neutron scattering \cite{Guitteny2013,Fennell2014,Ruminy2016,Ruminy2019,Turrini2021} and THz spectroscopy measurements \cite{Constable2017,Amelin2020} have revealed the presence of several magneto-elastic modes. In particular, it was shown  that a  so-called vibronic process \cite{Thalmeier1984} develops below 50~K and mixes the ground and first excited Tb$^{3+}$ crystal field doublets with acoustic phonons \cite{Constable2017}. A second vibron, visible below 200~K, couples the first excited doublet with a silent optical phonon mode \cite{Constable2017}. 

\begin{figure}
\includegraphics[width=1\columnwidth]{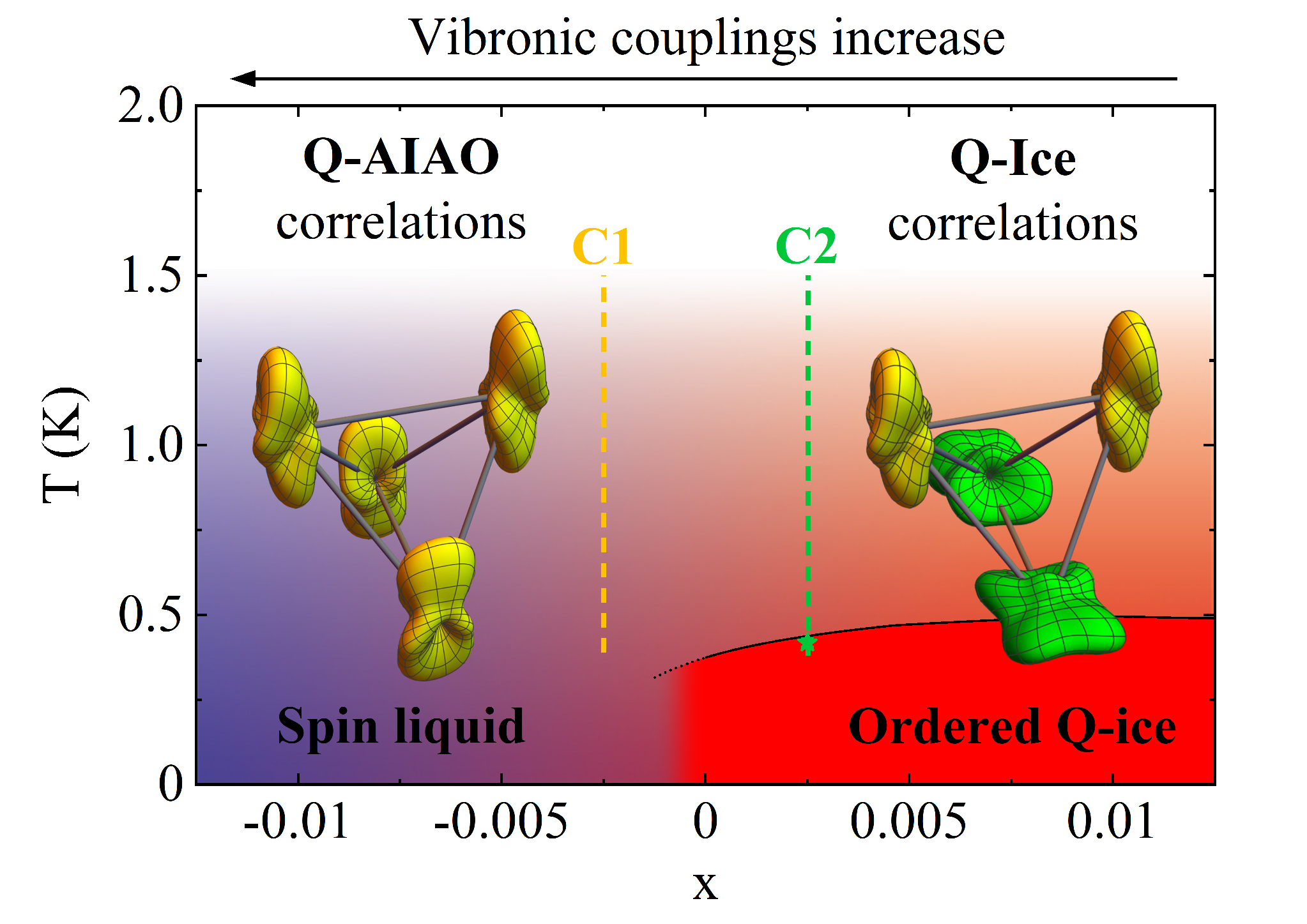}
\caption{Tb$_{2+x}$Ti$_{2-x}$O$_{7+y}$ phase diagram as a function of the temperature and the off stoechimetry $x$ deduced from this work and from Refs \cite{Taniguchi2013,Wakita2016}. Quadrupolar all-in-all-out like correlations are favored by vibronic couplings for $x\lesssim 0$ while quadrupolar ice correlations are promoted by multipolar interactions for $x\gtrsim 0$ leading to an ordered quadrupolar ice at low temperatures. Our two single crystals C1 and C2 have been characterized by specific heat measurements in the temperature range shown by the dashed lines. C2 orders below 0.42~K (green star). The configuration of the four quadrupoles on one tetrahedron in both regions is displayed.}
\label{fig:Fig1}
\end{figure}

In this paper, we used synchrotron based THz spectroscopy to provide evidence for a competition between the previously described vibronic processes and the low-temperature quadrupolar order of Tb$_{2+x}$Ti$_{2-x}$O$_{7+y}$ (see Fig. \ref{fig:Fig1}). Our measurements reveal that the spin lattice couplings are stronger when $x<0$ than $x>0$. We then show that these couplings promote quadrupolar all-in-all-out like correlations (Q-AIAO), which compete with the quadrupolar ice like correlations (Q-Ice) that prevails for $x\gtrsim 0$, where an ordered quadrupolar ice is finally stabilized at very low temperature. The spin liquid state of Tb$_{2+x}$Ti$_{2-x}$O$_{7+y}$ compounds with $x\lesssim0$ is therefore caused by a subtle equilibrium between magnetic interactions and the competition between these two types of quadrupolar correlations.

\section{Experimental results \& discussion}

\subsection{THz measurments}

High resolution THz spectroscopy measurements were performed at Synchrotron SOLEIL on the AILES beamline from 200~K to 6~K in the 8 $-$ 24~cm$^{-1}$ range (resolution 0.2~cm$^{-1}$) on a C2 crystal, previously used in a magneto-optical study \cite{Amelin2020}. We measured two different plaquettes cut out of this crystal, perpendicular to a $\left<111\right>$ and $\left<110\right>$ direction of the cubic pyrochlore lattice (thickness 150~$\mu$m / 170~$\mu$m, diameter 1.7~mm / 1.2~mm, respectively). The experimental conditions were the same as the high-resolution THz study previously done on another crystal C1 \cite{Constable2017}, whose results are shown in the present paper to allow a direct comparison between the two crystals. To determine their position in the phase diagram as a function of the off stoichiometry, they were both characterized by specific heat measurements in the temperature range 10 $-$ 400~K owing to a physical property measurement system (PPMS) with a ${}^{3}$He insert, using the same samples. THz absorption spectra were calculated solving Maxwell equations with the magnetic susceptibility tensor deduced from eigenenergies and eigenvectors of the Hamiltonian \cite{Amelin2020,Alexanian2021}. Absorption peaks were modeled by the same pseudo-Voigt function $0.7 L + 0.3 G$, where $L$ and $G$ are Lorenzian and Gaussian functions with the same width at half maximum chosen equal to 2.7~cm$^{-1}$, in agreement with the observed line width of Ref. \cite{Constable2017} on the C1 sample.


\begin{figure}
\includegraphics[width=1\columnwidth]{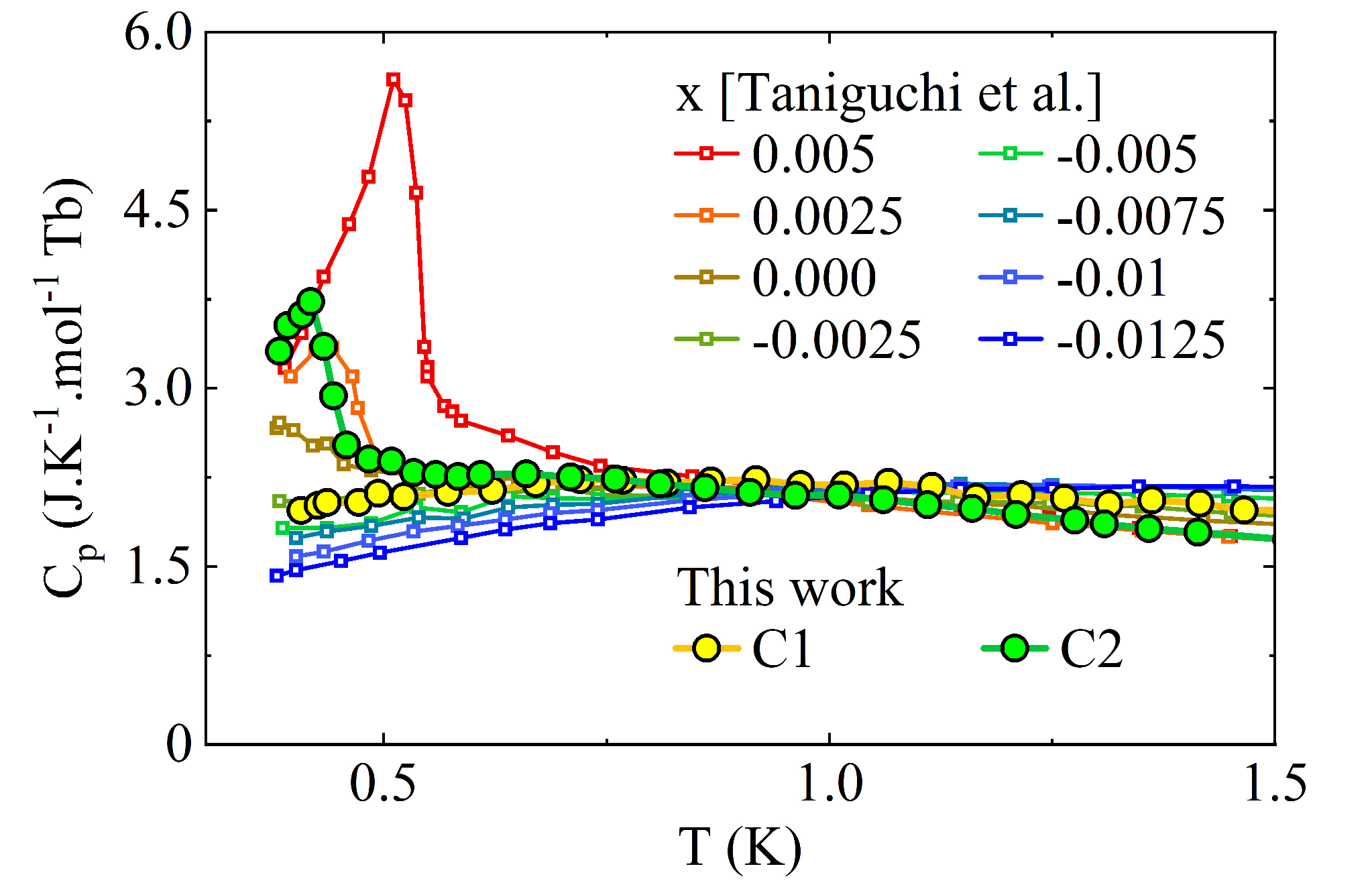}
\caption{Tb$_{2+x}$Ti$_{2-x}$O$_{7+y}$ specific heat as a function of temperature for different off-stoichiometry $x$. Solid symbols: this study for crystals C1 (yellow) and C2 (green). Open symbols: data from Ref. \cite{Taniguchi2013}. The comparison of both data sets allows us to deduce that $x\approx-0.0025$ for C1 and $x\approx+0.0025$ for C2.}
\label{fig:Fig2}
\end{figure}
The specific heat characterization for the two crystals is given in Fig. \ref{fig:Fig2} and, to evaluate their off stoichiometry, compared to measurements on samples with different $x$-values from Ref. \cite{Taniguchi2013}. A clear peak is visible for C2 at $T_{\mathrm{c}} =$ 0.42~K, which yields $x\approx+0.0025$. This peak reveals the onset of a quadrupolar order. Quite differently, no peak is observed in C1, confirming that this crystal, whose off-stoichiometry is estimated to $x\approx-0.0025$, presents a quadrupolar order but remains in a spin liquid state \cite{Petit2012}.

\begin{figure*}
\includegraphics[width=1\textwidth]{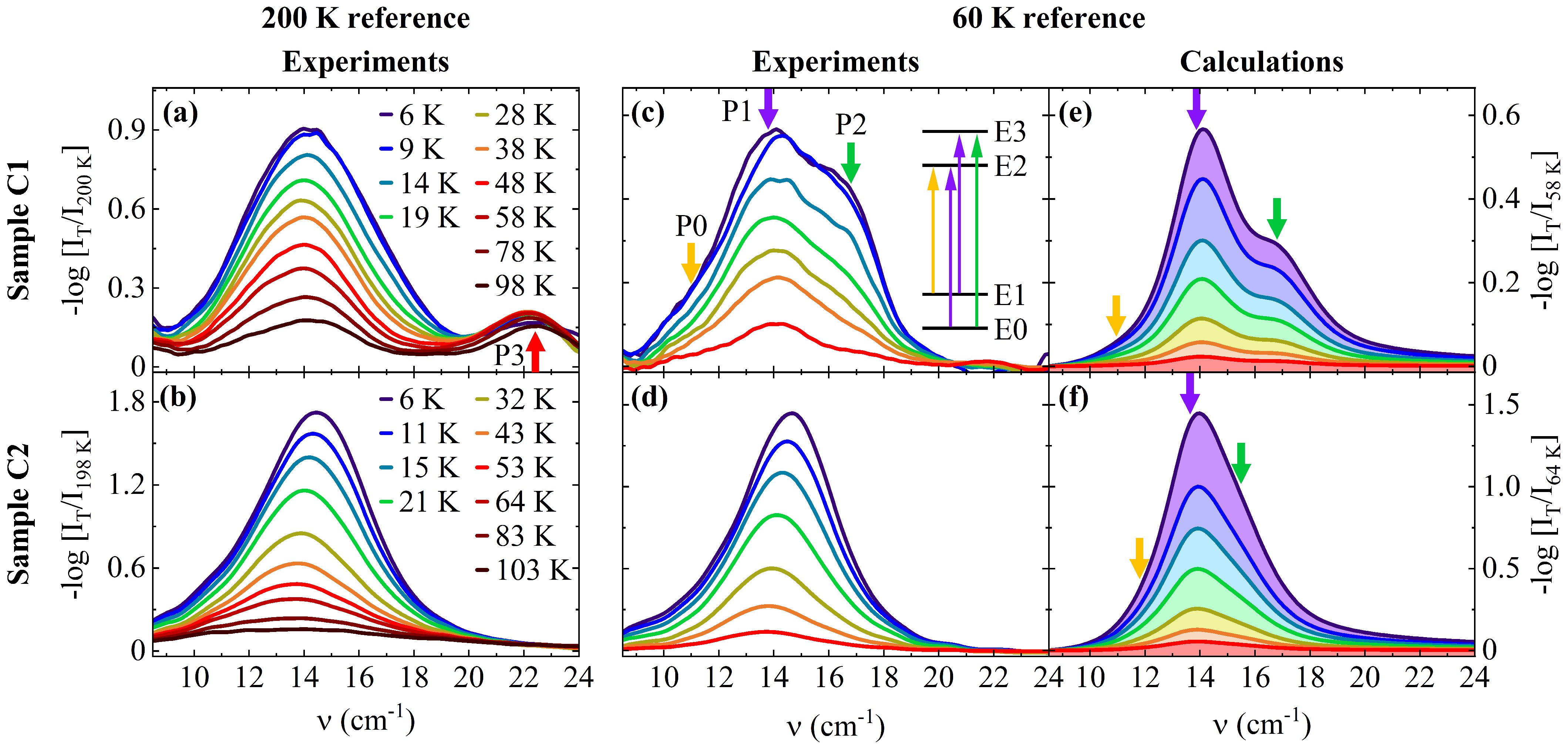}
\caption{Tb$_{2+x}$Ti$_{2-x}$O$_{7+y}$ THz spectra as a function of temperature for crystal C1 (upper panels) and C2 (lower panels). These differential absorptions are calculated by dividing the sample spectrum of interest by one at a higher temperature chosen to emphasize different processes: (a), (b) with a high temperature reference $T\approx $200~K, and (c), (d) with a low temperature reference $T\approx $60~K. Calculations of the same quantity with the low temperature reference for a vibronic parameter (e) $D = -$0.38~meV and (f) $D = -$0.24~meV (f). The THz wave vector was $\bm{k}\parallel\dca*{111}$ and its polarization $\bm{e}\parallel\dca*{0\bar{1}1}$, $\bm{h}\parallel\dca*{2\bar{1}\bar{1}}$. The measured and calculated positions of the P0 $-$ P3 peaks are shown by the arrows. In the inset of (c), the splitting by the vibronic coupling of the two ground and first excited doublets is schematized.}
\label{fig:Fig3}
\end{figure*}

The THz spectroscopy results on crystal C2 are presented in Fig. \ref{fig:Fig3} for different temperatures, together with those previously obtained on crystal C1. In Figs. \ref{fig:Fig3}(a) and \ref{fig:Fig3}(b), the reference spectrum is taken at 200~K. The broad absorption peak, centered around 14~cm$^{-1}$, is observed in both samples and is assigned to the first Tb$^{3+}$ crystal field excited doublet. Another smaller peak centered around 22~cm$^{-1}$ is present in the C1 spectra only, called P3 in Ref. \cite{Constable2017}, is experimental evidence of the high temperature vibronic process: While coupling with the first crystal field excited level, the transverse optical phonon, initially infrared inactive, becomes visible. Results on C2 reveal that P3 is absent, showing that this magnetoelastic process seems not to take place.

To highlight the off-stoichiometry impact on the low-temperature vibronic process, Figs. \ref{fig:Fig3}(c) and \ref{fig:Fig3}(d) show the same spectra but with a 60~K reference, a significantly lower temperature where the high-temperature vibronic process observed in C1 becomes temperature independent. The previous results for crystal C1 that are reproduced here show a fine structure: Besides the central peak P1 at 13.8~cm$^{-1}$, another one, P2, is observed at 16.8~cm$^{-1}$. Further analysis \cite{Constable2017} shows that a third peak, P0, is also present at 11~cm$^{-1}$. These three peaks were associated to a splitting of the ground and first crystal-field excited doublets due to the low-temperature vibronic process that involves acoustic phonons. Quite surprisingly, for C2, we only observe one single peak. Note also that spectra with different wave polarizations or wave vectors are similar in C2 whereas they vary in C1 (see Supplemental Material \cite{SupMat}). It is intriguing that the fine structure of C1 is not visible in C2 all the more so that the influence of the low-temperature vibronic process was observed in the magneto-optical study on this sample: It was a key ingredient to understand the magnetic field dependence of its THz spectra, in particular the splitting of one of the field-induced transition as reported in Ref. \cite{Amelin2020}.

\subsection{THz spectra calculation}

To understand these evolutions of the THz spectra, numerical simulations have been carried out using the vibronic model developed previously for crystal C1 \cite{Constable2017, Amelin2020}. Then the Hamiltonian of our simulations contains two parts: the crystal-field Hamiltonian $\widehat{\mathcal{H}}_{\mathrm{cf}}$ and the low temperature vibronic coupling one $\widehat{\mathcal{H}}_{\mathrm{vib}}$. Using the local threefold axis of each Tb$^{3+}$ site (point group symmetry $\mathcal{D}_{\mrm{3d}}$) as the quantification $z$ axis, they can be written in the same form for the four Tb$^{3+}$ ions at the vertices of each tetrahedron that constitutes the pyrochlore lattice. In this case, $\widehat{\mathcal{H}}_{\mathrm{cf}}$ writes
\begin{equation}
\begin{aligned}
\widehat{\mathcal{H}}_{\mathrm{cf}} &= \theta_{2}\lambda_{2}^{0}B_{0}^{2}\widehat{\mathcal{O}}_{2}^{0} + \theta_{4}\dpa*{\lambda_{4}^{0}B_{0}^{4}\widehat{\mathcal{O}}_{4}^{0} + \lambda_{4}^{3}B_{3}^{4}\widehat{\mathcal{O}}_{4}^{3}}\\
&+ \theta_{6}\dpa*{\lambda_{6}^{0}B_{0}^{6}\widehat{\mathcal{O}}_{6}^{0} + \lambda_{6}^{3}B_{3}^{6}\widehat{\mathcal{O}}_{6}^{3} + \lambda_{6}^{6}B_{6}^{6}\widehat{\mathcal{O}}_{6}^{6}}
\label{eq:Hcf}
\end{aligned}
\end{equation}
where $\widehat{\mathcal{O}}_{k}^{q}$ are Stevens operators, $B_{q}^{k}$ are crystal fields parameters, $\theta_{k}$ are reduced matrix elements, and $\lambda_{k}^{q}$ are numerical coefficients (see Supplemental Material \cite{SupMat}). For $\widehat{\mathcal{H}}_{\mathrm{vib}}$, we used the Hamiltonian first proposed by Thalmeier and Fulde \cite{Thalmeier1982} and developed for Tb$_{2}$Ti$_{2}$O$_{7}$ in Refs.\cite{Constable2017, Alexanian2021}, in its simplest form with a single vibronic coupling parameter $D$, assumed to be the only $x$ dependent parameter in our calculations:
\begin{equation}
\widehat{\mathcal{H}}_{\mathrm{vib}} = \theta_{2}\lambda_{2}^{1}D\left(\widehat{\mathcal{O}}_{2}^{1} + \widehat{\mathcal{O}}_{2}^{-1}\right)\textrm{.}
\label{eq:Hvib_simpl}
\end{equation}
This effective Hamiltonian is identical to the one describing a static strain. More information  can be found in the Supplemental Material \cite{SupMat}. Due to this vibronic process, both the ground and first excited doublets are almost equally split producing three distinct energy levels $E_{1} - E_{3}$ [see insert of Fig. \ref{fig:Fig3}(c)] obtained in the calculation by diagonalizing $\widehat{\mathcal{H}}_{\mathrm{cf}}+\widehat{\mathcal{H}}_{\mathrm{vib}}$ (results given in Table \ref{tab:energies_cef_Hme}). For C1, we find good agreement between numerical simulations and experimental results with $D = -$0.38~meV [see Figs. \ref{fig:Fig3}(c), and \ref{fig:Fig3}(e)]. The calculated peak positions $E_{2}-E_{1} =$ 10.9~cm$^{-1} = \mrm{P}0$, $E_{2}-E_{0} =$ 13.8~cm$^{-1} \approx \mrm{P}1$,  $E_{3}-E_{1} =$ 13.9~cm$^{-1} \approx \mrm{P}1$, and $E_{3}-E_{0} =$ 16.8~cm$^{-1} = \mrm{P}2$ are in remarkable agreement with experimental ones, $\mrm{P}0 =$ 11.0~cm$^{-1}$, $\mrm{P}1 =$ 13.8~cm$^{-1}$ and $\mrm{P}2 =$ 16.8~cm$^{-1}$. Furthermore, our calculations reproduce qualitatively the experimental spectral shape with two clearly visible peaks (P1 and P2) and another one only slightly visible P0. For C2, since no visible signature of the vibronic process is seen, we used the magneto-optical measurements of Ref. \cite{Amelin2020} (reproduced in the Supplemental Material \cite{SupMat}) to fix the vibronic parameter $D = -$0.24~meV. Using this coupling, the C2 spectrum shape is also well reproduced with one single broad excitation, as observed, embedding the three non-resolved peaks at $E_{2}-E_{1} =$ 11.8~cm$^{-1} = \mrm{P}0$, $E_{2}-E_{0} =$ 13.6~cm$^{-1}\approx \mrm{P}1$,  $E_{3}-E_{1} =$ 13.7~cm$^{-1}  \approx \mrm{P}1$, and $E_{3}-E_{0} =$ 15.5~cm$^{-1} = \mrm{P}2$ [see Fig. \ref{fig:Fig3}(f)].
\begin{table}
\begin{ruledtabular}
\begin{tabular} {lccc}
\noalign{\vskip 0.5mm}
& $E_{1}$ & $E_{2}$ & $E_{3}$  \\
\hline\noalign{\vskip 0.5mm} 
C1 ($D=-$0.38~meV) & $2.85/0.35$ & $13.8/1.71$ & $16.8/2.08$ \\
C2 ($D=-$0.24~meV) & $1.83/0.23$ & $13.6/1.69$ & $15.5/1.92$ \\
\end{tabular}
\end{ruledtabular}
\caption{Calculated energies of the three first excited levels (in cm$^{-1}$/meV) of C1 and C2.}
\label{tab:energies_cef_Hme}
\end{table}

We note however that the experimental low-temperature dependence is not well captured in our single parameter model. This might be explained by the influence, below 10~K, of interactions between spins and quadrupole degrees of freedom. Indeed, they are necessary for the onset of quadrupole order below 0.5~K and are also required to understand the observed low-temperature magnetic correlations \cite{Kadowaki2018,Kadowaki2019,Kadowaki2022}. Nevertheless, our THz observations as well as our vibronic interpretation are rather compatible with published data using optical spectroscopy \cite{Lummen2008,Zhang2021} and inelastic neutron scattering \cite{Taniguchi2013,Kadowaki2018,Ruminy2019}. See details in the Supplemental Material \cite{SupMat}. This being said, our model allows us to understand the differences observed between both samples: The three peaks P0, P1 and P2 are present in both cases and due to vibronic processes but are not experimentally resolved for C2 because of a lower vibronic coupling.

\subsection{Pseudo spin approach for the x dependent phase diagram}

To go beyond and study the implication of the $x$-dependence of the vibronic coupling on the ground state of Tb$_{2+x}$Ti$_{2-x}$O$_{7+y}$, it is useful to adopt a pseudo-spin description where the vibronic Hamiltonian acts as a perturbation on the crystal field ground state doublet. In this approach,
\begin{equation}
\widehat{\mathcal{H}}_{\mathrm{vib}}^{(\mrm{p.s.})} \propto \theta_{2}\lambda_{2}^{1}D(\widehat{\sigma}^{x}-\widehat{\sigma}^{y})
\end{equation}
where $\widehat{\sigma}^{x,y,z}$ are Pauli matrices (see Supplemental Material \cite{SupMat}). Note that for a non Kramers ion as Tb$^{3+}$, the longitudinal ($z$) pseudospin component of the doublet is related to the magnetic dipolar degree of freedom and can couple to a magnetic field while the transverse ($x$, $y$) components represent electric quadrupoles degrees of freedom \cite{Onoda2010,Lee2012,Curnoe2018} that are non magnetic. Since the vibronic Hamiltonian has the same form for the four Tb$^{3+}$ ions of a tetrahedron, it will favor an orientation of the transverse components of all the Tb$^{3+}$ ions along the same local direction. This will result in planar ferropseudospin also called Q-AIAO configurations, which may eventually order at low temperature. On the other hand, it has been shown that quadrupolar interactions between Tb$^{3+}$ ions favor a planar antiferropseudospin order \cite{Takatsu2016,Kadowaki2022}, which corresponds to an ordered ice of transverse pseudo-spin components \cite{Petit2016,Yan2017,Rau2019}. The corresponding Tb$^{3+}$ electronic densities calculated \cite{Schmitt1986,Kusunose2008}  for both Q-AIAO and Q-Ice configurations are represented in Fig. 1 (see Supplemental Material \cite{SupMat}) with distinct colors for the two different quadrupolar states. The picture which emerges in Tb$_{2+x}$Ti$_{2-x}$O$_{7+y}$ is thus the following: while Tb$-$Tb interactions are strong enough to stabilize a long-range ordered quadrupolar ice for $x\gtrsim0$, they compete with the low-temperature vibronic process Which promotes all-in-all-out like quadrupole correlations. As a consequence, the long range order is suppressed for $x\lesssim0$, namely when magnetoelastic couplings are sufficiently strong as observed  for crystal C1. For larger negative $x$ values, with possibly even stronger vibronic couplings, we then foresee the existence of an ordered Q-AIAO phase that would be interesting to investigate.

At this stage, it is interesting to point out possible effects of disorder. First, random strains associated to defects caused by the off-stoichiometry have not been considered. An approach similar to the one developed in Ref. \cite{Boldyrev2022} would be interesting to develop. The Hamiltonian may have the form of Eq. \ref{eq:Hvib_simpl} (see Supplemental Material), the $D$ parameter no longer being uniform. Second, as in other non-Kramer pyrochlores \cite{Petit2016,Wen2017,Martin2017}, disorder may produce competition between the quadrupolar order and a quantum spin liquid \cite{Savary2017}. Such a scenario cannot apply to Tb$_{2+x}$Ti$_{2-x}$O$_{7+y}$ since the splitting of the crystal-field levels are weaker in crystal C1 presenting a quadrupolar state, than in C2 with a spin-liquid state, in contradiction with the phase diagram proposed in Ref. \cite{Benton2018}. It is also worth noting that in Tb$_2$Ti$_2$O$_7$ the first excited doublet should also be taken into account. However, it seems likely that it will not have a significant impact on our interpretation. On one hand, as Eq. \ref{eq:Hvib_simpl} shows, the vibronic process will still favor the same orientation of the quadrupoles on a whole tetrahedron. On the second hand, it will probably only slightly modify the phase boundaries and position of Tb$_{2+x}$Ti$_{2-x}$O$_{7+y}$ in the phase diagram since the energy of the first excited crystal-field level is about an order of magnitude larger than the energy scale of typical interactions in rare-earth pyrochlores \cite{Rau2019}. One more question we can address is the origin of the strong $x$ dependence of vibronic couplings. While a quantitative answer is beyond the scope of this study, the very low value of $x$ for which changes occur suggests that this dependence has a collective origin. One can therefore reasonably think that vibronic processes might be strongly affected by local defects recalling that a vibronic ground state actually results from a hybridization between local degrees of freedom (crystal-field excitations) and collective degrees of freedom (acoustic phonons).\\

\section{Conclusion}

In summary, using high-resolution synchrotron-based THz spectroscopy coupled with quantum calculations, we show that vibronic processes, namely hybridizations between crystal-field levels and phonons, are strongly affected by a small change of stoichiometry $x$ in Tb$_{2+x}$Ti$_{2-x}$O$_{7+y}$. Our measurements reveal that the amplitude of these two processes (one appearing below 200~K and one below 50~K) are greater when $x<0$ than $x>0$. All spectra measured in two different samples, at different temperatures, and with magnetic fields (results of Ref. \cite{Amelin2020}) are qualitatively reproduced by our calculations which use a single, sample-dependent, vibronic coupling parameter. We show that these vibronic couplings affect the ground state in the sense that they promote quadrupolar correlations antagonistic to those resulting from quadrupolar interactions and leading to an order at very low temperature for $x\gtrsim0$. This quadrupolar order is even destroyed for $x\lesssim0$ when these vibronic couplings become stronger, and the subtle admixture between magnetic interactions and these two quadrupolar effects in competition leaves a spin liquid ground state. This scenario is not without similarities with frustrated spin systems where correlations of magnetic dipoles can compete producing unconventional behaviors, in particular the destabilization of the spin-ice state when spin-ice and all-in-all-out correlations coexist \cite{BrooksBartlett2014}. In the present case, the very rich behavior of Tb$_{2+x}$Ti$_{2-x}$O$_{7+y}$ can be interpreted on the basis of such a frustrating mechanism but based on higher rank multipoles, opening the way to more general cases of multipolar frustration, beyond the case of pyrochlore materials \cite{Hattori2016,Tsunetsugu2021}.\\

\begin{acknowledgments}
We thank J\'er\^ome Debray for the shaping and orientation of the crystals and Pierre Lachkar for the specific heat measurements. E.C. acknowledges financial support by the Austrian Science Fund (grant no. P32404-N27).
\end{acknowledgments}

\end{document}


\title{Supplemental material for “Vibronic collapse of ordered quadrupolar ice in the pyrochore magnet Tb$_{2+x}$Ti$_{2-x}$O$_{7+y}$"}

\author {Y. Alexanian}\email{yann.alexanian@neel.cnrs.fr}\altaffiliation{Present adress: Institut Laue Langevin, 38 000 Grenoble, France}
\affiliation{Universit\'e Grenoble Alpes, CNRS, Institut N\'eel, 38000 Grenoble, France}
\author{J. Robert}
\affiliation{Universit\'e Grenoble Alpes, CNRS, Institut N\'eel, 38000 Grenoble, France}
\author{V. Simonet}
\affiliation{Universit\'e Grenoble Alpes, CNRS, Institut N\'eel, 38000 Grenoble, France}
\author {B. Langérôme}
\affiliation{Synchrotron SOLEIL, L’Orme des Merisiers, 91192 Gif-sur-Yvette, France}
\author {J.-B. Brubach}
\affiliation{Synchrotron SOLEIL, L’Orme des Merisiers, 91192 Gif-sur-Yvette, France}
\author {P. Roy}
\affiliation{Synchrotron SOLEIL, L’Orme des Merisiers, 91192 Gif-sur-Yvette, France}
\author{C. Decorse}
\affiliation{ ICMMO, Universit\'e Paris-Saclay, CNRS, 91400 Orsay, France}
\author{E. Lhotel}
\affiliation{Universit\'e Grenoble Alpes, CNRS, Institut N\'eel, 38000 Grenoble, France}
\author{E. Constable}
\affiliation{Institute of Solid State Physics, TU Wien, 1040 Vienna, Austria}
\affiliation{Universit\'e Grenoble Alpes, CNRS, Institut N\'eel, 38000 Grenoble, France}
\author {R. Ballou}
\author {S. De Brion}\email{sophie.debrion@neel.cnrs.fr}
\affiliation{Universit\'e Grenoble Alpes, CNRS, Institut N\'eel, 38000 Grenoble, France}

\date{\today}

\maketitle

\section{Measurements with other polarizations}
\begin{figure}
\includegraphics[width=1\columnwidth]{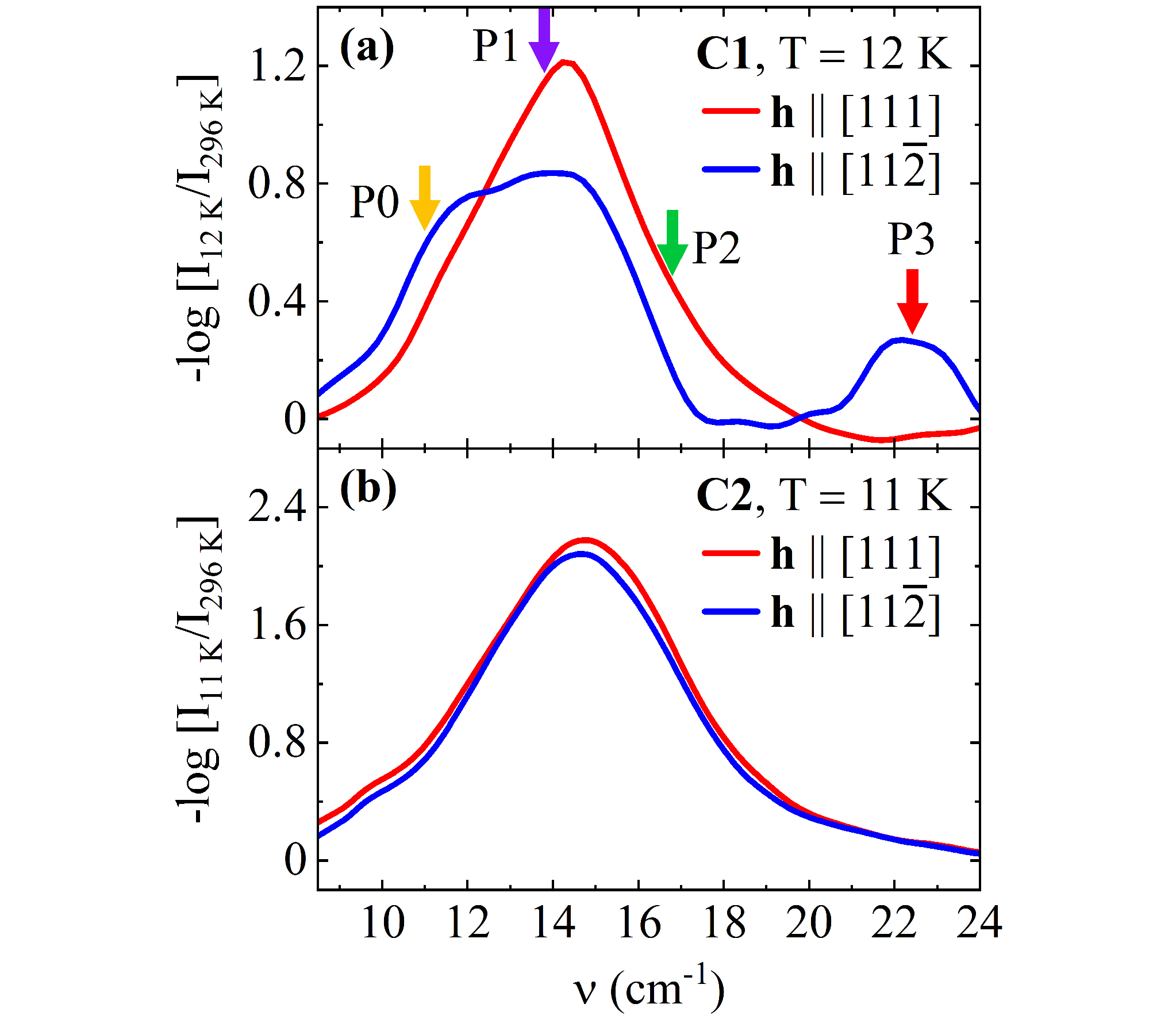}
\caption{THz spectra measured at 12~K and analyzed with a temperature reference $T\approx$300~K for $\bm{k}\parallel\left[0\bar{1}1\right]$ and two wave polarizations, $\bm{h}\parallel\left[111\right]$ (in red) and $\bm{h}\parallel\left[11\bar{2}\right]$ (in blue) on (a) C1 and (b) C2. The measured and positions of the P0 to P3 peaks are shown by the arrows.}
\label{fig:FigA}
\end{figure}

The  THz spectra measured with other THz polarizations using samples with  $\bm{k}\parallel\left[1\bar{1}0\right]$ confirm the scenario described for $\bm{k}\parallel\left[111\right]$, that is the presence of a sample dependent vibronic coupling. Fig. \ref{fig:FigA} reports the experimental data obtained at approximately 10~K with two different polarizations: $\bm{e}\parallel\left[11\bar{2}\right]$, $\bm{h}\parallel\left[111\right]$ and $\bm{e}\parallel\left[\bar{1}\bar{1}\bar{1}\right]$, $\bm{h}\parallel\left[11\bar{2}\right]$ for C1 (Fig. \ref{fig:FigA}a) and C2 (Fig. \ref{fig:FigA}b) with a temperature reference around 300~K. Again, the behavior is very different between the two single crystals. For C1, a single peak is visible for $\bm{h}\parallel\left[111\right]$ while a significantly complex spectrum is observed with $\bm{h}\parallel\left[11\bar{2}\right]$: the intensity of the central peak (P1) is lower compared to $\bm{h}\parallel\left[111\right]$ and two new peaks appear, centered at 11~cm$^{-1}$ (P0) and 22~cm$^{-1}$ (P3). The presence of this fine structure, together with the polarization dependence are the signature of vibronic couplings, see Ref. \cite{Constable2017, Amelin2020} for details. Conversely, wave polarization has no effect on C2 spectra: only the central peak is visible and its intensity is independent of the $\bm{h}$ direction, showing that the effects of vibronic couplings are not visible for this sample and are therefore weaker.

\section{Hamiltonian diagonalization and wavefunctions}

The crystal field hamiltonian is defined in the main text by
\begin{equation}
\begin{aligned}
\widehat{\mathcal{H}}_{\mathrm{cf}} &= \theta_{2}\lambda_{2}^{0}B_{0}^{2}\widehat{\mathcal{O}}_{2}^{0} + \theta_{4}\dpa*{\lambda_{4}^{0}B_{0}^{4}\widehat{\mathcal{O}}_{4}^{0} + \lambda_{4}^{3}B_{3}^{4}\widehat{\mathcal{O}}_{4}^{3}}\\
&+ \theta_{6}\dpa*{\lambda_{6}^{0}B_{0}^{6}\widehat{\mathcal{O}}_{6}^{0} + \lambda_{6}^{3}B_{3}^{6}\widehat{\mathcal{O}}_{6}^{3} + \lambda_{6}^{6}B_{6}^{6}\widehat{\mathcal{O}}_{6}^{6}},
\label{eq:Hcf}
\end{aligned}
\end{equation}
where $\widehat{\mathcal{O}}_{k}^{q}$ are Stevens operators, $B_{k}^{q}$ are the  crystal field parameters deduced from experiments \cite{Amelin2020} (displayed in Table \ref{tab:cef_TbTi}), $\theta_{k}$ are reduced matrix elements \cite{Stevens1952,Hutchings1964} (reproduced in Table \ref{Reduced_matrix_element_Tb3+}) and $\lambda_{k}^{q}$ are numerical coefficients \cite{Danielsen1972} (given in Table \ref{lambda_parameters}). The wavefunctions for the ground state doublet $\ket{\psi_{\pm}^{0}}$ and the first excited doublet $\ket{\psi_{\pm}^{1}}$ obtained from the diagonalization of the crystal field hamiltonian $\widehat{\mathcal{H}}_{\mathrm{cf}}$ are given in Table \ref{Wavefunction_CF}. 
\begin{table}
\begin{ruledtabular}
\begin{tabular} {cccccc}
$B_{0}^{2}$ & $B_{0}^{4}$ & $B_{3}^{4}$ & $B_{0}^{6}$ & $B_{3}^{6}$ & $B_{6}^{6}$ \\
\hline\noalign{\vskip 1mm} 
$51.4$ & $295$ & $114$ & $64.6$ & $-84.2$ & $129$ \\
\end{tabular}
\end{ruledtabular}
\caption{Crystal field parameters of Tb$_{2+x}$Ti$_{2-x}$O$_{7+y}$ used for calculations  (in meV), taken from Ref. \cite{Amelin2020}.}
\label{tab:cef_TbTi}
\end{table}
\begin{table}[ht!]
\begin{ruledtabular}
\begin{tabular} {ccc}
$\theta_{2}$ & $\theta_{4}$ & $\theta_{6}$ \\
\hline  \noalign{\vskip 1mm}   
$-1/99$ & $2/16335$ & $-1/891891$ \\
\end{tabular}
\end{ruledtabular}
\caption{Reduced matrix element $\theta_{k}(J)$ for Tb$^{3+}$ ($J = 6$). From Refs. \cite{Stevens1952,Hutchings1964}.}
\label{Reduced_matrix_element_Tb3+}
\end{table}
\begin{table}[ht!]
\begin{ruledtabular}
\begin{tabular} {cccccc}
$\lambda_{2}^{0}$ & $\lambda_{4}^{0}$ & $\lambda_{4}^{3}$ & $\lambda_{6}^{0}$ & $\lambda_{6}^{3}$ & $\lambda_{6}^{6}$ \\
\hline \noalign{\vskip 1mm}
$1/2$ & $1/8$ & $\sqrt{35}/2$ & $1/16$ & $\sqrt{105}/8$ & $\sqrt{231}/16$ \\
\end{tabular}
\end{ruledtabular}
\caption{$\lambda_{k}^{q}$ numerical coefficients. Adapted from Ref. \cite{Danielsen1972}.}
\label{lambda_parameters}
\end{table}
\clearpage

\begin{table}
\begin{ruledtabular}
\begin{tabular} {ccccc}
& $\ket{\psi^{-}_{0}} (0.00)$ & $\ket{\psi^{+}_{0}} (0.00)$ & $\ket{\psi^{-}_{1}} (13.5)$ & $\ket{\psi^{+}_{1}} (13.5)$ \\
\hline\noalign{\vskip 1mm}   
$\ket{+6}$ & & & & \\
$\ket{+5}$ & & $-0.345$ & & $0.888$ \\
$\ket{+4}$ & $-0.912$ & & $-0.370$ & \\
$\ket{+3}$ & & & & \\
$\ket{+2}$ & & $0.184$ & & $-0.246$\\
$\ket{+1}$ & $0.125$ & & $0.114$ &  \\
$\ket{\;\;\: 0}$ & & & & \\
$\ket{-1}$ & & $0.125$ & & $0.114$ \\
$\ket{-2}$ & $-0.184$ & & $0.246$ & \\
$\ket{-3}$ & & & & \\
$\ket{-4}$ & & $0.912$ & & $0.370$ \\
$\ket{-5}$ & $-0.345$ & & $0.888$ & \\
$\ket{-6}$ & & & & \\
\end{tabular}
\end{ruledtabular}
\caption{Wavefunction of the $4$ lowest lying levels, obtained by diagonalization of the crystal field hamiltonian (Eq. \ref{eq:Hcf}). The value in brackets after the name of a wavefunction is its associated eigenenergy (in cm$^{-1}$).}
\label{Wavefunction_CF}
\end{table}


The vibronic hamiltonian is defined in the main text as 
\begin{equation}
\widehat{\mathcal{H}}_{\mathrm{vib}} = \theta_{2}\lambda_{2}^{1}D\left(\widehat{\mathcal{O}}_{2}^{1} + \widehat{\mathcal{O}}_{2}^{-1}\right)
\label{eq:Hvib_simpl}
\end{equation}
with $\theta_{2}=-1/99$, $\lambda_{2}^{1}=\sqrt{6}$ and $D$ the vibronic coupling parameter. It has the same form as the one describing a static strain:  shear is described by the quadrupolar operator $\widehat{\mathcal{O}}_{2}^{1}$ and $\widehat{\mathcal{O}}_{2}^{-1}$ in the local $\mathcal{D}_{\mathrm{3d}}$ point group while compression is described by $\widehat{\mathcal{O}}_{2}^{0}$. This last will not produce any splitting of the Eg levels since this operator is already present in the crystal field Hamiltonian. However, no static deformation has been observed in  Tb$_2$Ti$_2$O$_7$, and  the hamiltonian \ref{eq:Hvib_simpl} rather describes the vibronic process (typically a dynamical modulation of the oxygen cages around each Tb). This is an effective hamiltonian likely to account for the effects of vibrons in the zero-wave vector limit, close to the zone center where the excitations are experimentally probed, and since acoustic phonons are involved, in the quasi-static limit as well. This quasi-static limit is justified considering the different time scales involved. Indeed, the THz technique is able to probe the instantaneous deformation produced by the phonon if the time scale of the measurement is shorter or comparable to the one of the phononic process. Since our THz measurements are not limited by the instrumental resolution, the appropriate time scale is the one given by the energy and life time of the excitations. They are both in the 1$-$10~cm$^{-1}$ range, which corresponds to a time scale around $10^{-11}$~s. On the other hand, an experiment with a longer time scale will probe the average, non-distorted environment. This is the case for ultrasound measurements as reported for Tb$_{2}$Ti$_{2}$O$_{7}$ \cite{Nakanishi2011}. There, no elastic deformation has been observed for a frequency range 5$-$30~MHz, i.e. a time scale of the order of  $10^{-7}$~s. A dynamical approach as the one performed by Thalmeier and Fulde \cite{Thalmeier1982} would have been of use in order to characterize the dispersion effects associated to the vibrons. It was however not necessary to develop it here since THz experiments are essentially zone center techniques probing the excitations in the zero-wave vector limit.

Using this effective vibronic hamiltonian, together with the crystal field hamiltionian, we can derive the wavefunctions of the first four levels (denoted $\ket{\varphi_{0}}$, $\ket{\varphi_{1}}$, $\ket{\varphi_{2}}$ and $\ket{\varphi_{3}}$) obtained from the diagonalization of $\widehat{\mathcal{H}}_{\mathrm{cf}}+ \widehat{\mathcal{H}}_{\mathrm{vib}}$ with $D=-$0.38~meV (sample C1) and with $D=-$0.24~meV (sample C2) are given in Tables \ref{Wavefunction_CF_Vib_C1} and \ref{Wavefunction_CF_Vib_C2} respectively. Note that $\ket{\varphi_{0,1}}$ are mainly built from $\ket{\psi^{\pm}_{0}}$ while $\ket{\varphi_{2,3}}$ are mainly built from $\ket{\psi^{\pm}_{1}}$.

\begin{table*}
\begin{ruledtabular}
\begin{tabular} {ccccc}
$D=-$0.38~meV & $\ket{\varphi_{0}} (0.00)$ & $\ket{\varphi_{1}} (2.85)$ & $\ket{\varphi_{2}} (13.8)$ & $\ket{\varphi_{3}} (16.6)$ \\
\hline\noalign{\vskip 1mm} 
$\ket{+6}$ & & & & \\
$\ket{+5}$ & $-0.301$ & $-0.092-0.119\mrm{i}$ & $-0.142+0.641\mrm{i}$ & $0.128+0.589\mrm{i}$ \\
$\ket{+4}$ & $0.445+0.428\mrm{i}$ & $0.083-0.670\mrm{i}$ & $0.140-0.091\mrm{i}$ & $-0.170+0.271\mrm{i}$ \\
$\ket{+3}$ & $-0.010\mrm{i}$ & & & $0.017$ \\
$\ket{+2}$ & $0.144$ & $0.064+0.083\mrm{i}$ & $0.041-0.186\mrm{i}$ & $-0.034-0.158\mrm{i}$ \\
$\ket{+1}$ & $-0.058-0.056\mrm{i}$ & $-0.012+0.098\mrm{i}$ & $-0.056+0.037\mrm{i}$ & $0.047-0.075\mrm{i}$ \\
$\ket{\;\;\: 0}$ & & & & \\
$\ket{-1}$ & $0.081$ & $0.060+0.078\mrm{i}$ & $-0.015+0.066\mrm{i}$ & $0.019+0.087\mrm{i}$ \\
$\ket{-2}$ & $0.104+0.100\mrm{i}$ & $0.013-0.104\mrm{i}$ & $-0.160+0.104\mrm{i}$ & $0.086-0.136\mrm{i}$ \\
$\ket{-3}$ & & $0.012$ & & $0.017$ \\
$\ket{-4}$ & $0.618$ & $0.413+0.535\mrm{i}$ & $-0.036+0.163\mrm{i}$ & $0.068+0.312\mrm{i}$ \\
$\ket{-5}$ & $0.217+0.209\mrm{i}$ & $0.019-0.150\mrm{i}$ & $-0.550+0.358\mrm{i}$ & $0.321-0.510\mrm{i}$ \\
$\ket{-6}$ & & & & \\
\end{tabular}
\end{ruledtabular}
\caption{Wavefunction of the $4$ lowest lying levels, obtained by diagonalization of the crystal field hamiltonian (Eq. \ref{eq:Hcf}) and the vibronic coupling hamiltonian (Eq. \ref{eq:Hvib_simpl}) with a parameter $D=-$0.38~meV. The value in brackets after the name of a wavefunction is its associated eigenenergy (in cm$^{-1}$). Only coefficients $>10^{-2}$ are shown.}
\label{Wavefunction_CF_Vib_C1}
\end{table*}

\begin{table*}
\begin{ruledtabular}
\begin{tabular} {ccccc}
$D=-$0.24~meV & $\ket{\varphi_{0}} (0.00)$ & $\ket{\varphi_{1}} (1.84)$ & $\ket{\varphi_{2}} (13.6)$ & $\ket{\varphi_{3}} (15.5)$ \\
\hline\noalign{\vskip 1mm}   
$\ket{+6}$ & & & & \\
$\ket{+5}$ & $0.282+0.025\mrm{i}$ & $0.108+0.157\mrm{i}$ & $-0.163+0.626\mrm{i}$ & $-0.126-0.598\mrm{i}$ \\
$\ket{+4}$ & $-0.404-0.479\mrm{i}$ & $-0.120+0.653\mrm{i}$ & $0.178-0.106\mrm{i}$ & $0.163-0.254\mrm{i}$ \\
$\ket{+3}$ & & & & $-0.011$ \\
$\ket{+2}$ & $-0.140-0.012\mrm{i}$ & $-0.065-0.095\mrm{i}$ & $0.046-0.178\mrm{i}$ & $0.034+0.162\mrm{i}$ \\
$\ket{+1}$ & $0.054+0.064\mrm{i}$ & $0.017-0.093\mrm{i}$ & $-0.063+0.037\mrm{i}$ & $-0.047+0.072\mrm{i}$ \\
$\ket{\;\;\: 0}$ & & & & \\
$\ket{-1}$ & $-0.083$ & $-0.054-0.078\mrm{i}$ & $-0.018+0.070\mrm{i}$ & $-0.018-0.084\mrm{i}$ \\
$\ket{-2}$ & $-0.090-0.107\mrm{i}$ & $-0.021+0.114\mrm{i}$ & $-0.158+0.094\mrm{i}$ & $-0.089+0.139\mrm{i}$ \\
$\ket{-3}$ & & & & $-0.011$ \\
$\ket{-4}$ & $-0.625-0.054\mrm{i}$ & $-0.375-0.548\mrm{i}$ & $-0.052+0.200\mrm{i}$ & $-0.062-0.295\mrm{i}$ \\
$\ket{-5}$ & $-0.183-0.217\mrm{i}$ & $-0.035+0.187\mrm{i}$ & $-0.555+0.331\mrm{i}$ & $-0.331+0.514\mrm{i}$ \\
$\ket{-6}$ & & & & \\
\end{tabular}
\end{ruledtabular}
\caption{Wavefunction of the $4$ lowest lying levels, obtained by diagonalization of the crystal field hamiltonian (Eq. \ref{eq:Hcf}) and the vibronic coupling hamiltonian (Eq. \ref{eq:Hvib_simpl}) with a parameter $D=-$0.24~meV. The value in brackets after the name of a wavefunction is its associated eigenenergy (in cm$^{-1}$). Only coefficients $>10^{-2}$ are shown.}
\label{Wavefunction_CF_Vib_C2}
\end{table*}

\section{THz absorption calculations with magnetic field}

\begin{figure*}[ht!]
\includegraphics[width=1\textwidth]{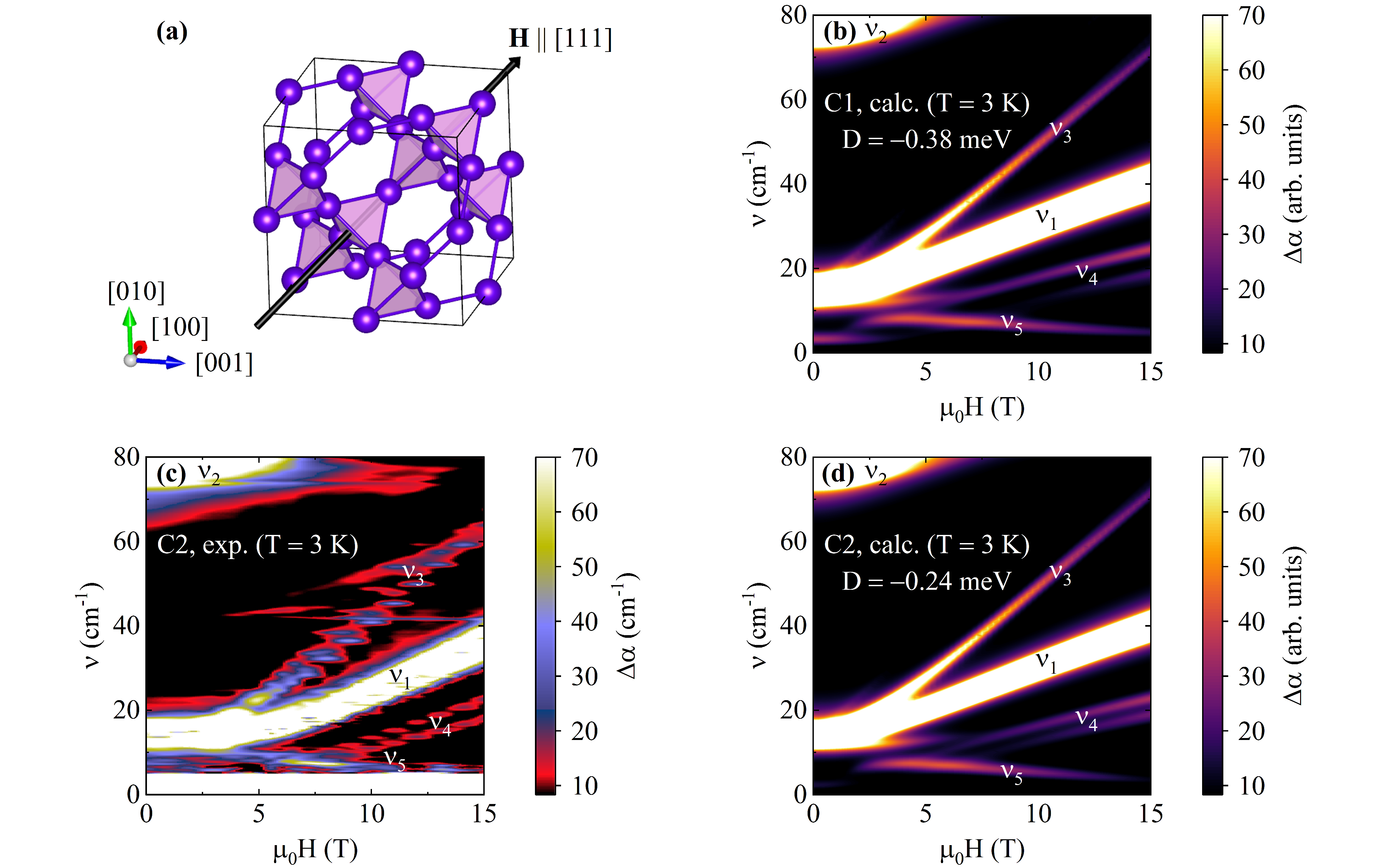}
\caption{THz spectra at 3~K when the magnetic field is applied along the [111] direction. (a) Sketch of the pyrochlore structure with the [111] direction \cite{Momma2011}. (b) Calculated spectra for sample C1 with a vibronic coupling $D=-$0.38~meV. (c) Measured THz spectra for sample C2 reproduced from Ref. \cite{Amelin2020}. (d) Calculated spectra for sample C2 with a vibronic coupling $D=-$0.24~meV. }
\label{fig:FigB}
\end{figure*}

\begin{table}
\begin{ruledtabular}
\begin{tabular} {ccccc}
Site & $1$ & $2$ & $3$ & $4$ \\
\hline\noalign{\vskip 1mm}  
$\sqrt{6}~\bm{x}_{i}$ & $(\bar{2},1,1)$ & $(\bar{2},\bar{1},\bar{1})$ & $(2,1,\bar{1})$ & $(2,\bar{1},1)$ \\
$\sqrt{2}~\bm{y}_{i}$ & $(0,\bar{1},1)$ & $(0,1,\bar{1})$ & $(0,\bar{1},\bar{1})$ & $(0,1,1)$ \\
$\sqrt{3}~\bm{z}_{i}$ & $(1,1,1)$ & $(1,\bar{1},\bar{1})$ & $(\bar{1},1,\bar{1})$ & $(\bar{1},\bar{1},1)$ \\
\end{tabular}
\end{ruledtabular}
\caption{Local basis of the $4$ Tb$^{3+}$ ions of a tetrahedron.}
\label{tab:local_axis}
\end{table}

Our calculations at 3~K when the magnetic field is applied along the [111] direction of the pyrochlore lattice are reported in Fig. \ref{fig:FigB} for the two samples. For crystal C1, we use a vibronic parameter $D=-$0.38~meV deduced from the zero field THz data where the splitting of the ground and first excited crystal field levels is clearly visible. For crystal C2, no splitting is detectable in zero field but the signature of the vibronic coupling is observed in the magnetic field dependent spectra reproduced in Fig. \ref{fig:FigB}(b). A vibronic coupling parameter $D=-$0.24~meV is able to account for all the observed features, in particular the splitting of the lower, weak branch ($\nu_{4}$) detected above 10~T. Details can be found in Ref. \cite{Amelin2020}.

We comment on different conventions between the calculations of Ref. \cite{Amelin2020} and the one presented in this work. First, the $D$ parameters used here are renormalized by the coefficients  $\theta_{2}$ and $\lambda_{2}^{1}$ (see equation (2) in the main text) giving rise to a factor of the order of $2.5\times 10^{-3}$ with respect to Ref. \cite{Amelin2020}. Second, the $B_{4}^{3}$ and $B_{6}^{3}$ crystal field parameters both change sign between the present work and Ref. \cite{Amelin2020}. This has no impact on observables in zero magnetic field but also with a magnetic field if combined to an appropriate change of local frame for the 4 sites on a tetrahedron. In order to use the conventional sign of $B_{4}^{3}$ and $B_{6}^{3}$ usually reported in the recent literature, we used for this work the local axes defined in Table \ref{tab:local_axis} (following the convention used in Ref. \cite{Yan2017}) different from the one used in Ref. \cite{Amelin2020}. This actually yields the same good agreement between the experimental and calculated magnetic field dependence of THz spectra.

Time resolved THz absorption spectra measurements with an applied magnetic field $\bv{H}\parallel\dca*{111}$ up to 6.9~T at $T=$ 1.6~K were also reported in Ref. \cite{Zhang2021}. The experimental results are in agreement with our measurements although the authors are not able to observe the splitting of the $\nu_{4}$ branch at these values of magnetic field. To reproduce their data, they performed calculations ignoring the hybridization between the crystal field levels and the acoustic phonons (the vibronic coupling we use) and instead including other spin lattice couplings in the form of a magnetic-field tunable crystal-field environment that lowers the Tb$^{3+}$ local symmetry from $\mathcal{D}_{\mrm{3d}}$ to $\mathcal{C}_{1}$. However, one should notice that, for such a symmetry lowering, either the oxygen or the terbium ions should move under the influence of the magnetic field. It is difficult to ensure that such effects will induce larger static displacements than the dynamical ones promoted by phonons, especially for low magnetic field. Moreover, this interpretation cannot capture our zero magnetic field spectra where vibronic couplings are already observed, mainly in C1. We believe then that vibronic coupling is the main effect responsible for the discrepancies between the measured THz spectra in magnetic field and the calculations when only the crystal field and the Zeeman terms in the hamiltonian are considered.

\section{Comparison with other optical and inelastic neutron scattering measurements}

The THz results presented in this study show that the excitation around  13~cm$^{-1}$ (1.5~meV), is split into three subpeaks and that the value of this splitting depends on the off stoichiometry (it decreases when $x$ increases). In our interpretation, these three peaks result from the transition between the ground state and the first excited doublets which are both split due to a magneto-vibrational hybridization. It is important to note first that our synchrotron based  THz measurements are not limited by the energy resolution which is very small (typically 0.2~cm$^{-1}$). This is due to the high intensity and stability of the synchrotron light in the 8$-$50~cm$^{-1}$ range which competes nicely with laboratory sources such as in Ref. \cite{Zhang2021}. Raman spectroscopy at such low energy is not easy either because of the presence of the central elastic peak at zero energy. Indeed, the Raman study presented in Ref. \cite{Lummen2008} mostly consider energies above 50~cm$^{-1}$. 

This is different with inelastic neutron scattering since, depending on the choice of the wavelength and instrument (triple-axis and time-of-flight), it can provide a sufficient energy resolution in the good energy window to observe the features we identified in our THz study. In Ref. \cite{Ruminy2019} for instance, a splitting of the 1.5~meV excitation into three subpeaks is reported, in agreement with the THz data presented in Fig. 3 of the main article. It is found that, at very low temperature (0.7~K), these peaks disperse weakly (compared to ordered systems). No significant differences were observed in two crystals with different off-stoechiometry ($x=0.0013$ and $x=-0.007$). However, these measurements probe different points in the reciprocal space than our zone-center THz measurements, and are at temperatures were the amplitude of the dispersion ($\le$ 0.5~meV) is higher than the splitting we observe. Concerning the splitting of the ground state, a small splitting of 0.1~meV was reported at 0.1~K \cite{Taniguchi2013,Kadowaki2018}. In these studies, different crystals with different well-controlled stoichiometry were used. This 0.1~meV level becomes narrower and slightly dispersive below the quadrupolar order transition temperature for $x>0$, while it remains broad and weaker for $x<0$. This level is less observable above 1~K which is consistent with the fact that it should be populated above this temperature such that the transition between the two split ground state levels should not be detectable. This splitting is a bit smaller than what we propose (0.35~meV for C1 and 0.23~meV for C2). However, we note that the hybridization with an acoustic phonon, steeply dispersive, is most influent close to the zone center. There, in these neutron data, the excitation seems to be pushed upward in energy and this is where our THz measurements are most sensitive compared to neutron scattering. It is very likely then that it can be compatible with our experimental results and interpretation.

\section{Pseudo-spin model and order parameter}

We recall that for Tb$^{3+}$ ions ($J=6$) in a $\mathcal{D}_{3\mathrm{d}}$ point group symmetry, the crystal field splits the ground multiplet into 5 singlets and 4 doublets. In Tb$_{2}$Ti$_{2}$O$_{7}$, experiments show that both the ground and first excited energy levels are doublets. Their associated wave functions are of the form
\begin{equation}
\ket{\psi^{\pm}_{l}} = \alpha_{5}^{l}\ket{\pm5} \pm \alpha_{2}^{l}\ket{\pm 2} + \alpha_{-1}^{l}\ket{\mp 1} \pm \alpha_{-4}^{l}\ket{\mp 4}\textrm{.}
\label{eq:wavefnc}
\end{equation}
which is the case of the calculated wavefunctions of Table \ref{Wavefunction_CF}. The projected expression of a multipolar operator $\wh{\mathcal{O}}_{k}^{q}$ on the ground state doublet $l=0$ is then
\begin{equation}
\widehat{\mathcal{P}}\wh{\mathcal{O}}_{k}^{q}\widehat{\mathcal{P}}^{-1} = 
\begin{pmatrix}
\mel{\psi^{+}_{0}}{\wh{\mathcal{O}}_{k}^{q}}{\psi^{+}_{0}} & \mel{\psi^{+}_{0}}{\wh{\mathcal{O}}_{k}^{q}}{\psi^{-}_{0}}\\
\mel{\psi^{-}_{0}}{\wh{\mathcal{O}}_{k}^{q}}{\psi^{+}_{0}} & \mel{\psi^{-}_{0}}{\wh{\mathcal{O}}_{k}^{q}}{\psi^{-}_{0}}
\end{pmatrix}
\label{eq:pseudo_spin}
\end{equation}
with $\widehat{\mathcal{P}} = \ket{\psi^{+}_{0}}\bra{\psi^{+}_{0}} + \ket{\psi^{-}_{0}}\bra{\psi^{-}_{0}}$ the projection operator.

\noindent $\widehat{\mathcal{P}}\wh{\mathcal{O}}_{k}^{q}\widehat{\mathcal{P}}^{-1}$  being a $2\times2$ matrix, it can be expressed as a linear combination of the Pauli matrices $\wh{\sigma}^{x,y,z}$ (or $\wh{\sigma}^{\pm,z}$) and the identity matrix $\wh{I}$:
\begin{equation}
\widehat{\mathcal{P}}\wh{\mathcal{O}}_{k}^{q}\widehat{\mathcal{P}}^{-1} = \sum_{\alpha}C_{k,q}^{\alpha}\wh{\sigma}^{\alpha} + C^{I}\wh{I}.
\end{equation}
In this equation, $C_{k,q}^{\alpha}$ and $C^{I}$ are numerical coefficients depending only on the ground state doublet wavefunction coefficients $\alpha^{0}_{n}$. For instance, one has
\begin{equation}
\begin{aligned}
\widehat{\mathcal{P}}\wh{\mathcal{O}}_{2}^{1}\widehat{\mathcal{P}}^{-1} = 
C_{2,1}^{x}\begin{pmatrix}
0 & 1\\
1 & 0
\end{pmatrix} = C_{2,1}^{x}\wh{\sigma}^{x}\\
\widehat{\mathcal{P}}\wh{\mathcal{O}}_{2}^{-1}\widehat{\mathcal{P}}^{-1} = 
C_{2,-1}^{y}\begin{pmatrix}
0 & -i\\
i & 0
\end{pmatrix} = C_{2,-1}^{y}\wh{\sigma}^{y}
\end{aligned}
\end{equation}
with $C_{2,1}^{x} = -C_{2,-1}^{y} \equiv C = -6\sqrt{\frac{5}{2}}\alpha_{2}^{0}\alpha_{-1}^{0} + 9\sqrt{\frac{11}{2}}\alpha_{5}^{0}\alpha_{-4}^{0}$. With the coefficients of Table \ref{Wavefunction_CF}, $C\approx -6.86$. The projected form of the vibronic coupling hamiltonian (Eq. \ref{eq:Hvib_simpl} and Eq. (2) of the main text) writes in this approach
\begin{equation}
\widehat{\mathcal{P}}\widehat{\mathcal{H}}_{\mathrm{vib}}\widehat{\mathcal{P}}^{-1}\equiv \widehat{\mathcal{H}}_{\mathrm{vib}}^{(\mrm{p.s.})} = \theta_{2}\lambda_{2}^{1}DC(\widehat{\sigma}^{x}-\widehat{\sigma}^{y})
\label{eq:vibrons}
\end{equation}
which corresponds to Eq. (3) of the main text. 

At this point, it is important to stress that, in the case of non-Kramers ions, the projection of odd rank multipolar operators  will only involve the $z$ component of the pseudo-spin while the projection of even rank multipolar operator will only involve $x$ and $y$ components of the pseudo-spin and the identity. This is due to the fact that $\wh{\sigma}^{z}=\ket{\psi^{+}_{0}}\bra{\psi^{+}_{0}}- \ket{\psi^{-}_{0}}\bra{\psi^{-}_{0}}$ transforms as a time reversal odd quantity (like a magnetic dipole) while $\wh{\sigma}^{x}=\ket{\psi^{+}_{0}}\bra{\psi^{-}_{0}} + \ket{\psi^{-}_{0}}\bra{\psi^{+}_{0}}$ and $\wh{\sigma}^{y}=-\mrm{i}\ket{\psi^{+}_{0}}\bra{\psi^{-}_{0}} +\mrm{i}\ket{\psi^{-}_{0}}\bra{\psi^{+}_{0}}$ transforms as time reversal even quantities (like an electric quadrupole) by the symmetry operations of the local point group $\mathcal{D}_{3\mathrm{d}}$ \cite{Onoda2010,Curnoe2018}.

\begin{table*}
\begin{center}
\begin{tabular} {lc}
\hline
\hline\noalign{\vskip 0.5mm}
Order & Order parameter \\
\hline\noalign{\vskip 0.5mm}
$\Gamma_{3}$ (AIAO) &
$m_{\Gamma_{3}} = \Psi_{1} = \mathcal{S}^{z}_{1}+\mathcal{S}^{z}_{2}+\mathcal{S}^{z}_{3}+\mathcal{S}^{z}_{4}$\\
\noalign{\vskip 3mm}  
$\Gamma_{5}$ &
$\bv{m}_{\Gamma_{5}} = \begin{pmatrix}
\Psi_{2}\\
\Psi_{3}\\
\end{pmatrix} = 
\begin{pmatrix}
\mathcal{S}^{x}_{1}+\mathcal{S}^{x}_{2}+\mathcal{S}^{x}_{3}+\mathcal{S}^{x}_{4}\\
\mathcal{S}^{y}_{1}+\mathcal{S}^{y}_{2}+\mathcal{S}^{y}_{3}+\mathcal{S}^{y}_{4}\\
\end{pmatrix}$ \\
\noalign{\vskip 3mm}
$\Gamma_{7}$ &
$\bv{m}_{\Gamma_{7}} = \begin{pmatrix}
\Psi_{4}\\
\Psi_{5}\\
\Psi_{6}\\
\end{pmatrix} = 
\begin{pmatrix}
\mathcal{S}^{y}_{1} + \mathcal{S}^{y}_{2} - \mathcal{S}^{y}_{3} - \mathcal{S}^{y}_{4}\\
-\sqrt{3}\dca*{\mathcal{S}^{x}_{1}-\mathcal{S}^{x}_{2}+\mathcal{S}^{x}_{3}-\mathcal{S}^{x}_{4}}/2 - \dca*{\mathcal{S}^{y}_{1}-\mathcal{S}^{y}_{2}+\mathcal{S}^{y}_{3}-\mathcal{S}^{y}_{4}}/2\\
+\sqrt{3}\dca*{\mathcal{S}^{x}_{1}-\mathcal{S}^{x}_{2}-\mathcal{S}^{x}_{3}+\mathcal{S}^{x}_{4}}/2 - \dca*{\mathcal{S}^{y}_{1}-\mathcal{S}^{y}_{2}-\mathcal{S}^{y}_{3}+\mathcal{S}^{y}_{4}}/2\\
\end{pmatrix}$ \\
\noalign{\vskip 3mm}
$\Gamma_{9}$ (OSI) &
$\bv{m}_{\Gamma_{9}^{\mathrm{A}}} = \begin{pmatrix}
\Psi_{7}\\
\Psi_{8}\\
\Psi_{9}\\
\end{pmatrix} = 
\begin{pmatrix}
\mathcal{S}^{z}_{1} + \mathcal{S}^{z}_{2} - \mathcal{S}^{z}_{3} - \mathcal{S}^{z}_{4}\\
\mathcal{S}^{z}_{1} - \mathcal{S}^{z}_{2} + \mathcal{S}^{z}_{3} - \mathcal{S}^{z}_{4}\\
\mathcal{S}^{z}_{1} - \mathcal{S}^{z}_{2} - \mathcal{S}^{z}_{3} + \mathcal{S}^{z}_{4}\\
\end{pmatrix}$ \\
\noalign{\vskip 3mm}
$\Gamma_{9}$ &
$\bv{m}_{\Gamma_{9}^{\mathrm{B}}} = \begin{pmatrix}
\Psi_{10}\\
\Psi_{11}\\
\Psi_{12}\\
\end{pmatrix} = 
\begin{pmatrix}
\mathcal{S}^{x}_{1} + \mathcal{S}^{x}_{2} - \mathcal{S}^{x}_{3} - \mathcal{S}^{x}_{4}\\
\dca*{\mathcal{S}^{x}_{1}-\mathcal{S}^{x}_{2}+\mathcal{S}^{x}_{3}-\mathcal{S}^{x}_{4}}/2 + \sqrt{3}\dca*{\mathcal{S}^{y}_{1}-\mathcal{S}^{y}_{2}+\mathcal{S}^{y}_{3}-\mathcal{S}^{y}_{4}}/2\\
\dca*{\mathcal{S}^{x}_{1}-\mathcal{S}^{x}_{2}-\mathcal{S}^{x}_{3}+\mathcal{S}^{x}_{4}}/2 - \sqrt{3}\dca*{\mathcal{S}^{y}_{1}-\mathcal{S}^{y}_{2}-\mathcal{S}^{y}_{3}+\mathcal{S}^{y}_{4}}/2\\
\end{pmatrix}$\\
\noalign{\vskip 0.5mm}
\hline
\hline
\end{tabular}
\end{center}
\caption{Pseudo-spin classical order parameter $\bv{k}\bv{=}\bv{0}$ for non-Kramers ions (adapted from Refs. \cite{Yan2017,Rau2019}). $\mathcal{S}^{\alpha}_{i}$ corresponds to the mean value of the pseudo-spin component $\alpha=x,y,z$ of the $i$ ion on a tetrahedron : $\mathcal{S}^{\alpha}_{i} \equiv \left<\wh{\sigma}^{\alpha}_{i}\right>$. $\bv{m}_{\Gamma_{5}}$, $\bv{m}_{\Gamma_{7}}$ and $\bv{m}_{\Gamma_{9}^{B}}$ order parameters involve transverse pseudo-spin components and the associated orders are therefore quadrupolar orders. Abreviations: AIAO= all-in all-out; OSI= ordered spin ice.}
\label{tab:order_params}
\end{table*}

This projection procedure is at the basis of a pseudo-spin approach for the interaction hamiltonian in pyrochlore systems. Different $i$, $j$  ions have to be considered now. Since  anisotropy axes are rotated from one site to another in a tetrahedron, it is necessary to distinguish the multipolar $\wh{\mathcal{O}}_{k}^{q}$ and the pseudo-spin $\wh{\sigma}^{\alpha}$ operators from one site to another. We denote $\wh{\mathcal{O}}_{k}^{q}\dpa{\wh{\bv{J}}_{i}}$ and $\wh{\sigma}^{\alpha}_{i}$ these operators for the $i$ ion. All possible multipolar interactions term $\wh{\mathcal{O}}_{k}^{q}\dpa{\wh{\bv{J}}_{i}}\mathcal{M}_{ij}^{k,q;k',q'}\wh{\mathcal{O}}_{k'}^{q'}\dpa{\wh{\bv{J}}_{j}}$ in the most general multipolar interaction hamiltonian $\wh{\mathcal{H}}_{\mrm{int}}$ (with $\mathcal{M}_{ij}^{k,q;k',q'}$ an element of the interaction parameter matrix $\mathcal{M}$) will be transposed into interactions between pseudo-spin \cite{Rau2019}. Restricting ourselves to first neighbour interactions, ommiting a constant term, and separating the Hamiltonian into a sum on the different tetrahedra $t$, the projection of the interaction hamiltonian will then be
\begin{equation}
\wh{\mathcal{P}}\wh{\mathcal{H}}_{\mrm{int}}\wh{\mathcal{P}}^{-1} = \sum_{t}\sum_{i,j\in t}{}^{t}\wh{\bv{\sigma}}_{i}\mathcal{J}_{ij}\dca*{t}\wh{\bv{\sigma}}_{j}
\label{eq:Hint}
\end{equation}
with $\mathcal{J}_{ij}\left[t\right]$ a $3\times 3$ matrix  depending on the bond $i-j$ of the tetrahedron $t$. This matrix encodes the numerous $\mathcal{M}_{ij}^{k,q;k',q'}$ possible parameters and the $C_{k,q}^{\alpha}$ coefficients.

If only classical (without intrication) ordered ground states are considered, it was shown that an energy minimum always exists at $\bv{k} = \bv{0} $ \cite{Yan2017}. In this case, it is sufficient to determine a configuration of vectors $\bv{\mathcal{S}}_{i} = \dba{\wh{\bv{\sigma}}_{i}}$ which minimizes the energy of a unique tetrahedron to find a ground state of the system. These ordered states at a tetrahedron scale are usually classified by the irreductible decomposition of the pseudo-spin configuration in the symmetry point group of the pyrochlore structure $\mathrm{m\bar{3}m}$. Table \ref{tab:order_params} gives the order parameters for the five distinct types of possible order in the case of non-Kramers doublets such as those of Tb$^{3+}$. Note that, since $\wh{\sigma}^{x,y}$ represents electric quadrupoles, any order involving these transverse pseudo-spin components ($\Gamma_{5}$, $\Gamma_{7}$ and one $\Gamma_{9}$) have to be interpreted as a quadrupolar order. 

The calculated phase diagram of Ref. \cite{Takatsu2016} shows two different quadrupolar orders denoted PF for ferropseudospin order and PAF for antiferropseudospin order. In Table \ref{tab:order_params}, PF order corresponds to $\Gamma_{5}$ while PAF order corresponds to either $\Gamma_{7}$ or $\Gamma_{9}$ quadrupolar order. In Ref. \cite{Kadowaki2022}, it is argued that the low temperature quadrupolar order of Tb$_{2+x}$Ti$_{2-x}$O$_{7}$ for $x\ge0$ corresponds to $\Gamma_{7}$. Adding our vibronic hamiltonian (Eq. \ref{eq:vibrons}) to the interaction hamiltonian (Eq. \ref{eq:Hint}) will not change the possible orders described in Table \ref{tab:order_params} since this does not depend on the detailed hamiltonian but only on the symmetry of the pyrochlore lattice and how the pseudo-spin components are transformed by these symmetries. It will amounts to apply a multiaxial all-in all-out transverse field in the pseudo-spin language since, with respect to the local symmetry axis, it acts the same way on each site of a tetrahedron. Consequently, the vibronic hamiltonian will tend to stabilize a transverse pseudo-spin order where all pseudo-spins are aligned the same way, the only possibility being the $\Gamma_{5}$ order described by the basis vector $\Psi_{2}$ and $\Psi_{3}$. This is seen formally when the vibronic hamiltonian $\wh{\mathcal{P}}\wh{\mathcal{H}}_{\mrm{vib}}\wh{\mathcal{P}}^{-1} \equiv \widehat{\mathcal{H}}_{\mathrm{vib}}^{(\mrm{p.s.})}$ is replaced by its classical limit $\mathcal{H}_{\mrm{vib}}$: the sum on the 4 sites of a tetrahedron gives
\begin{equation}
\sum_{i=1}^{4}\mathcal{H}_{\mrm{vib}}(i) \propto \sum_{i=1}^{4}\dpa*{\mathcal{S}_{i}^{x}-\mathcal{S}_{i}^{y}} = \dpa*{\Psi_{2}-\Psi_{3}}.
\end{equation}
Thus, similarely to the $z$ component of the pseudo-spin for which the spin ice ($\Gamma_{9}$) and the (spin) AIAO ($\Gamma_{3}$) configurations compete, this $\Gamma_{5}$ quadrupolar order competes with the $\Gamma_{7}$ order promoted by interactions.

\section{Angular dependence of the charge density}

In Fig. 1 of the main text, the angular dependence of the charge density is plotted using the wavefunctions $\ket{\varphi_{0}}$ and $\ket{\varphi_{1}}$ obtained for sample C2 ($D=-$0.24~meV, Table \ref{Wavefunction_CF_Vib_C2}). The Quadrupolar all-in all-out (Q-AIAO) state stabilized by the vibronic coupling corresponds to a charge density associated with the wavefunction $\ket{\varphi_{0}}$ on each site of the tetrahedron.  To represent the Quadrupolar Ice (Q-Ice) state, we plot the charge density associated with the wavefunction $\ket{\varphi_{0}}$ on two sites of the tetrahedron and the wavefunction $\ket{\varphi_{1}}$ on the two other sites since $\{\ket{\varphi_{0}},\ket{\varphi_{1}}\}$ depict the two opposite basis states of the pseudo-spin associated with the ground state doublet. We describe here the procedure allowing us to plot these angular dependences of the charge density  \cite{Schmitt1986,Kusunose2008}.

Let $\ket{\Psi}$ be an electronic state. Its wavefunction over spherical angles $(\theta,\phi)\equiv\Omega$ is determined through the projection $\Psi(\Omega) = \qsp{\Omega}{\Psi}$ where $\ket{\Omega}$ stands for a spherical angle state. One then deduces its angular probability distribution in the form
\begin{equation}
\mathcal{P}_{\Psi}(\Omega) = |\Psi(\Omega)|^{2} = \qsp{\Psi}{\Omega}\qsp{\Omega}{\Psi} = \Psi(\Omega)^{*}\Psi(\Omega).
\end{equation}
$\mathcal{P}_{\Psi}(\Omega)$, as any function of spherical angles, can be expanded over the basis of the spherical harmonics $Y_{k}^{q}(\Omega)$:
\begin{equation}
\mathcal{P}_{\Psi}(\Omega)=\sum_{k=0}^{+\infty}\sum_{q=-k}^{k}\zeta_{kq}(\Psi)Y_{k}^{q}(\Omega)
\label{eq:ang_proba}
\end{equation}
where $\zeta_{kq}$ are the coefficient of the expansion given by 
\begin{equation}
\zeta_{kq}(\Psi) =\int_{\mathcal{S}^{2}}\mrm{d}\Omega Y_{k}^{q*}(\Omega)\mathcal{P}_{\Psi}(\Omega).
\label{eq:coef}
\end{equation}

Spherical harmonics allow to define the spherical harmonics operators $\wh{\mathcal{Y}}_{k}^{q}$ by writing that their matrix elements on the spherical angle state space are written
\begin{equation}
\mel{\Omega}{\wh{\mathcal{Y}}_{k}^{q}}{\Omega'} = Y_{k}^{q}(\Omega)\delta(\Omega,\Omega').
\label{eq:spherical_harm_operators}
\end{equation}
The $\wh{\mathcal{Y}}_{k}^{q}$ are irreducible spherical tensor operators and the relation $Y_{k}^{q*}(\Omega) = (-1)^{q}Y_{k}^{-q}(\Omega)$ translate to $\wh{\mathcal{Y}}_{k}^{q\dagger} = (-1)^{q}\wh{\mathcal{Y}}_{k}^{-q}$. Rewriting Eq. \ref{eq:coef} so that Eq. \ref{eq:spherical_harm_operators} can be inserted within and using closure relation afterward allows to express $\zeta_{kq}(\Psi)$ coefficients in terms of matrix elements:
\begin{equation}
\zeta_{kq}(\Psi) = \mel{\Psi}{\wh{\mathcal{Y}}_{k}^{q\dagger}}{\Psi}.
\label{eq:matrix_elem}
\end{equation}

In our case, $\ket{\Psi}$ identifies to $\ket{\varphi_{n=0,1}}$ and belongs to the subspace of the ground multiplet ($J=6$ for Tb$^{3+}$). Note first that since the $\wh{\mathcal{Y}}_{k}^{q}$ operators are parity even for $k$ even and parity odd for $k$ odd (as the spherical harmonics), and that the states of the ground multiplet are of the same parity (since belonging to the same electronic configuration), the matrix elements of Eq. \ref{eq:matrix_elem} necessarily cancels out for odd $k$. For even $k$, they can be computed using the equivalent operator method \cite{Stevens1952} so that
\begin{equation}
\zeta_{kq}(\varphi_{n}) = \sqrt{\frac{2k+1}{4\pi}}\theta_{k}(J)\mel{\varphi_{n}}{\wh{\mathcal{R}}_{k}^{q\dagger}}{\varphi_{n}}
\label{eq:matrix_elem_Racah}
\end{equation}
where $\theta_{k}(J)$ are reduced matrix elements and $\wh{\mathcal{R}}_{k}^{q}$ are Racah operators (proportionnal to Wybourne operators $\wh{\mathcal{C}}_{k}^{q}$ by $\wh{\mathcal{C}}_{k}^{q} = \theta_{k}(J)\wh{\mathcal{R}}_{k}^{q}$). Let us also mention that since all $\theta_{k}$ coefficients are null for $k>6$ in the case of $f$ electrons, only the coefficents $\zeta_{kq}(\varphi_{n})$ with $k=0,2,4,6$ have to be calculated. The charge density of the wavefunction $\ket{\varphi_{n}}$ can then be obtain by inserting the coefficients of Eq. \ref{eq:matrix_elem_Racah} into Eq. \ref{eq:ang_proba}: it is the functions
$\mathcal{P}_{\varphi_{0}}(\Omega)$ and $\mathcal{P}_{\varphi_{1}}(\Omega)$ that are plotted in Fig. (1) of the main text. 

It could be more convenient to use the Stevens operators $\wh{\mathcal{O}}_{k}^{q}$ linked to Racah operators $\wh{\mathcal{R}}_{k}^{q}$ by
\begin{equation}
\wh{\mathcal{O}}_{k}^{q} = 
\begin{dcases}
\frac{1}{\lambda_{k}^{q}}\dpa*{\wh{\mathcal{R}}_{k}^{-\abs{q}}+(-1)^{q}\wh{\mathcal{R}}_{k}^{\abs{q}}}\textrm{ if $q > 0$,} \\
\frac{1}{\lambda_{k}^{q}}\wh{\mathcal{R}}_{k}^{q}\textrm{ if $q = 0$,}\\
\frac{\mathrm{i}}{\lambda_{k}^{q}}\dpa*{\wh{\mathcal{R}}_{k}^{-\abs{q}}-(-1)^{q}\wh{\mathcal{R}}_{k}^{\abs{q}}}\textrm{ if $q < 0$.}
\end{dcases}
\label{eq:Stevens_Racah}
\end{equation}
as spherical harmonics $Y_{k}^{q}(\Omega)$ are linked (up to a constant $\lambda_{k}^{q}/\sqrt{2}$) to tesseral (real) harmonics $Z_{k}^{q}(\Omega)$.
This gives
\begin{equation}
\begin{gathered}
\mathcal{P}_{\varphi_{n}}(\Omega)=\sum_{k=0}^{+\infty}\sum_{q=-k}^{k}\xi_{kq}(\varphi_{n})Z_{k}^{q}(\Omega),\\
\xi_{kq}(\varphi_{n}) = \frac{\lambda_{k}^{q}}{\sqrt{2}}\sqrt{\frac{2k+1}{4\pi}}\theta_{k}(J)\mel{\varphi_{n}}{\wh{\mathcal{O}}_{k}^{q}}{\varphi_{n}}.
\end{gathered}
\label{eq:matrix_elem_Stevens}
\end{equation}